\documentclass[journal]{IEEEtran}
\IEEEoverridecommandlockouts

\usepackage{silence}
\usepackage{cite}
\usepackage{amsmath}
\usepackage{algorithm,algorithmic}
\usepackage{graphicx}
\usepackage{subcaption}
\usepackage{dblfloatfix}
\usepackage{placeins}
\usepackage{booktabs}
\usepackage{multirow}
\usepackage{xcolor}
\usepackage{enumitem}
\usepackage{pifont}

\PassOptionsToPackage{hyphens}{url}
\usepackage{hyperref}
\hypersetup{hidelinks}
\newcommand{\cmark}{\ding{51}}
\newcommand{\xmark}{\ding{55}}
\newcommand{\circmark}{\scalebox{0.8}{$\bigcirc$}}
\newcommand{\gupair}[2]{#1\,$|$\,#2}

\newcommand{\N}{\mathbf{N}}
\newcommand{\R}{\mathbf{R}}
\newcommand{\V}{\mathcal{V}}
\newcommand{\E}{\mathcal{E}}
\newcommand{\D}{\mathcal{D}}
\newcommand{\B}{\mathcal{B}}
\newcommand{\G}{\mathcal{G}}

\graphicspath{{figures/}}

\begin{document}

\title{What Graph Unlearning Cannot Erase: Physics-Induced Leakage of Cyberattack Information in Electric Vehicle Charging Networks}

\author{Nanhong~Liu,~\IEEEmembership{Graduate Student Member,~IEEE,} Mucun Sun,~\IEEEmembership{Senior~Member,~IEEE,} and Jie~Zhang,~\IEEEmembership{Senior~Member,~IEEE}

    \thanks{Nanhong Liu is with the Department of Mechanical Engineering, The University of Texas at Dallas, Richardson, TX 75080, USA.}
    \thanks{Mucun Sun is with the Idaho National Laboratory, Idaho Falls, ID 83415, USA.}
    \thanks{Jie Zhang is with the Department of Mechanical Engineering and (Affiliated) Department of Electrical and Computer Engineering, The University of Texas at Dallas, Richardson, TX 75080, USA (e-mail: jiezhang@utdallas.edu).}
}

\maketitle

\begin{abstract}
Electric vehicle charging stations (EVCSs) can expose distribution feeders to cyberattacks.
While machine learning methods, including graph neural networks (GNNs), can localize which bus is compromised, significant challenges remain in data sharing and model training.
For example, data-sharing agreements grant EVCS operators the right to withdraw their charging power for privacy or security reasons from a deployed model, yet retraining from scratch on every request would repeat the full-retraining (FR) cost.
To address this, we study graph unlearning (GU) for EVCS cyberattack localization, formulated as a feature-level unlearning problem on a graph-level multi-label classification task.
We benchmark fine-tuning (FT) against two GU baselines on the IEEE 34-bus, 123-bus, and 8500-node distribution networks across three GNN backbones and cumulative unlearning scenarios.
Experiments show that a GNN unlearned or retrained without the withdrawn power still localizes cyberattacks at the forgotten EVCSs.
This physics-induced leakage floor arises because a cyberattack on an EVCS also shifts the voltages of electrically coupled buses.
The distribution system operator can thus promise an EVCS operator that the deployed GNN model no longer uses its charging power data, but the voltages still reveal cyberattacks at its EVCS.
\end{abstract}


\section{Introduction}
\label{sec:intro}

Electric vehicle charging stations (EVCSs) introduce new cyber-physical vulnerabilities at the grid edge of power distribution networks, where they operate alongside distributed energy resources and smart inverters~\cite{markovic2023machine}.
As electric vehicle (EV) adoption accelerates, these EVCSs become deeply embedded in distribution network operations and widen the attack surface.
Early work showed that manipulating the charging load of many EVs could threaten the frequency stability of the power grid~\cite{acharya2020public}.
Charge manipulation attacks (CMAs)~\cite{jahangir2024charge} manipulate the information exchanged during smart charging to shift the charging demand across the day.
In a distribution feeder, CMAs inject false charging profiles at the point of common coupling through false data injection or man-in-the-middle spoofing~\cite{jacob2025cyber}, and a single compromised bus can propagate voltage violations, load forecast errors, and cascading instability across feeders.
Cyberattack detection in power systems has been studied extensively, and a wide range of model-based and data-driven algorithms exist for detecting false data injection attacks~\cite{musleh2019survey, caleb2026false, olojede2025topology, uddin2025mp}.
For CMAs specifically, detection techniques ranging from statistical anomaly classifiers to deep learning models flag manipulated charging profiles from EVCS or market telemetry~\cite{jahangir2024charge, dubey2026cost}.
These techniques process each EVCS or session in isolation and do not localize the compromised buses within the feeder, which leaves the distribution system operator (DSO) without the bus-level evidence required to issue isolation, dispatch, or restoration commands.

A distribution feeder is naturally a graph, in which buses are nodes and lines are edges, and a manipulation at one EVCS perturbs the voltages of nearby buses through the power flow.
The evidence of a cyberattack is therefore spread over electrically coupled buses rather than confined to the compromised one, and a model that scores each bus in isolation discards the very signal that separates the attack source from its neighbors.
Graph neural networks (GNNs) capture this coupling by propagating measurements along the feeder topology~\cite{liao2021review}, \cite{sadric2026graph}, and GNN-based models have reached strong performance on cyberattack detection in power grids~\cite{boyaci2021graph, liu2026privacy} and on intrusion detection in the communication networks of EVCSs~\cite{poudel2026bidirectional}.
The same topology awareness also supports EVCS cyberattack localization, a step beyond detection.
We formulate it as a graph-level multi-label classification task in which each EVCS bus carries an independent label and the model output is the spatial attack pattern.
A GNN model learns from two kinds of data, the bus voltages that the DSO measures and the charging power that each EVCS operator shares.

Deletion obligations differ between the parties whose data enter such a model.
EV drivers are natural persons, and the ``right to be forgotten'' under the European Union's General Data Protection Regulation (GDPR)~\cite{mantelero2013eu} and the California Consumer Privacy Act (CCPA)~\cite{pardau2018ccpa} entitles them to request the removal of their personal charging data.
Charging power aggregated at a public EVCS is instead commercial data of the EVCS operator, a legal person.
The operator's right to withdraw it from a model trained by the DSO arises from data-sharing agreements.
The EU Data Act, which has applied since September 2025, lets the users of connected devices access, use, and share the data these devices generate and prohibits unfair contracts that prevent such sharing~\cite{ec2025dataact}.
In the energy sector, these devices include EV chargers~\cite{enode2025dataact}, so an EVCS operator can decide with whom the data of its chargers are shared.
Erasing the shared records, however, leaves their influence in a model already trained on them.
The EU Artificial Intelligence (AI) Act~\cite{eu2024aiact} adds a further layer, since it classifies AI systems used as safety components in the management and operation of electricity supply as high-risk and subjects their training data to data governance requirements.
A contractual withdrawal of this kind does not in general call for a formal deletion certificate, but the DSO still has to remove the influence of the withdrawn data, at a cost it can bear for every request.
This paper studies what such a removal achieves.
The removal alone does not establish legal compliance, which also depends on contractual terms and audit procedures.

Deletion requests also recur, since a feeder serves many EVCS operators and each request changes the data the model may use.
Retraining from scratch after every request removes the withdrawn data exactly but repeats the full training cost each time~\cite{shaik2024exploring}, so the DSO needs a cheaper way to update the deployed model.
A withdrawal, however, removes only the charging power.
The voltages remain under the control of the DSO, and a cyberattack at an EVCS changes the power flow and thus the voltages of electrically coupled buses.
The voltages and currents measured in a distribution grid are known to expose the load patterns of its customers~\cite{dvorkin2020differentially}.
Even a model retrained without the withdrawn records can therefore still reveal whether the EVCS was attacked.
The physics-induced cyberattack information leakage is the information on cyberattacks at an EVCS that the retained voltages still carry after its charging power is withdrawn.
The level that retraining leaves is the leakage floor.
It bounds what a DSO can promise an EVCS operator, because a model can forget the withdrawn charging power, but not its physical consequences, which the voltages still record.

Machine unlearning (MU)~\cite{cao2015towards} removes the influence of training data from a deployed model without full-retraining (FR), through exact or approximate unlearning~\cite{xue2026towards}.
Exact unlearning retrains only the sub-models whose data shards are affected, as in the SISA (Sharded, Isolated, Sliced, and Aggregated) scheme~\cite{bourtoule2021machine}.
Approximate unlearning instead estimates the parameter change, most often through the influence function~\cite{koh2017understanding}, a second-order correction built from the inverse Hessian.
The influence function has also been extended to features and labels, with first-order and second-order updates computed from the gradient difference between the original and the corrected training data, and the efficacy of these updates decreases as more features and labels are affected~\cite{warnecke2021machine}.
Fine-tuning (FT) the deployed model for a few epochs on the retained or corrected data is one of the simplest approximate schemes~\cite{jia2023model}, yet the influence of the deleted data can persist through it~\cite{ding2024understanding}.
In power and energy systems, MU has been applied to load forecasting~\cite{xu2024task}, to transformer fault localization~\cite{liu2026a}, and to the operation of EV charging reservations~\cite{lee2024learning}.
A recent survey~\cite{chahine2026survey} adds use cases such as smart meter data erasure and the removal of poisoned measurement data.
Physics-aware unlearning removes the influence of poisoned labels from the EVCS and photovoltaic inverter controllers of the IEEE 34-bus feeder~\cite{haider2026physics}.
That work shows that statistical unlearning alone does not restore physically safe operation.
An attacker can also misuse unlearning requests to make a deployed grid model forget what it has learned~\cite{paphitis2026machine}.

Graph unlearning (GU)~\cite{said2023survey} specializes MU to the nodes, edges, and features of a trained GNN.
GraphEraser~\cite{chen2022graph} partitions the graph for sharded retraining, at a cost in utility because partitioning breaks connectivity.
Graph influence function (GIF)~\cite{wu2023gif} adds multi-hop propagation to the influence function, IDEA (flexIble anD cErtified unleArning)~\cite{dong2024idea} certifies the correction for node, edge, and feature requests, the graph scattering transform (GST)~\cite{pan2023unlearning} certifies unlearning for graph classification, and GNNDelete~\cite{cheng2023gnndelete} learns layer-wise deletion operators.
\begin{table}[!b]
\centering
\caption{Unlearning methods implemented for graph classification in OpenGU~\cite{fan2026opengu}, combined with this work.}
\label{tab:gu_compare}
\footnotesize
\setlength{\tabcolsep}{5pt}
\begin{tabular*}{0.94\columnwidth}{@{\extracolsep{\fill}}lccccc@{}}
\toprule
Method & N-U & E-U & F-U & F(part)-U & Guarantee \\
\midrule
GraphEraser~\cite{chen2022graph}    & \gupair{--}{\cmark}     & \gupair{--}{\cmark}     & \gupair{--}{\cmark}     & \circmark & exact \\
GIF~\cite{wu2023gif}                & \gupair{\xmark}{\cmark} & \gupair{\xmark}{\cmark} & \gupair{\xmark}{\cmark} & \cmark & none \\
IDEA~\cite{dong2024idea}            & \gupair{\xmark}{\cmark} & \gupair{\xmark}{\cmark} & \gupair{\xmark}{\cmark} & \cmark & cert. \\
GST~\cite{pan2023unlearning}        & \gupair{\cmark}{\cmark} & \gupair{\cmark}{\cmark} & \gupair{\cmark}{\cmark} & \xmark & cert. \\
\midrule
FT                                  & \cmark & \cmark & \cmark & \cmark & none \\
\bottomrule
\end{tabular*}

\vspace{3pt}
\begin{minipage}{0.94\columnwidth}
\footnotesize
N-U, E-U, F-U: node, edge, and feature unlearning. F(part)-U: unlearn part of the features of the requested nodes.
For the existing methods, the N-U, E-U, and F-U entries follow Table~6 of OpenGU~\cite{fan2026opengu}, with the left symbol for the original code and the right one for OpenGU: \cmark\ implemented, \xmark\ not implemented, -- extension prevented by the design of the method.
The F(part)-U column and the FT row are assessed in this work: \cmark\ applicable, \circmark\ reduces to FR, \xmark\ not applicable.
Guarantee, what each method proves under its own assumptions: exact, the result matches retraining; cert., certified removal; none, no guarantee.
\end{minipage}
\end{table}
Fine-tuning the trained GNN forms another class of GU methods~\cite{zhang2025dynamic}, and when a GNN must forget manipulated training data, plain fine-tuning on the retained data comes close to retraining in some evaluations~\cite{kolipaka2024cognac}.
The OpenGU benchmark~\cite{fan2026opengu} integrates 16 GU algorithms on 37 datasets spanning social, biological, and citation networks.
Only four of them, GraphEraser, GIF, IDEA, and GST, are implemented for graph classification.
\autoref{tab:gu_compare} combines the graph classification entries of these four methods with the request studied here, which removes part of the features of the requested nodes in every training graph.
GIF and IDEA apply to this request and serve as the baselines.
GraphEraser reduces to FR, since such a request affects every shard.
GST does not apply, since it unlearns a linear classifier on fixed scattering features and its certification assumes that fewer nodes are removed at once than the smallest graph contains.
A more recent benchmark~\cite{jain2026is} evaluates GU methods for node classification against retraining at four levels, namely accuracy, prediction fidelity, logits, and parameters, and finds that retraining from scratch often remains the more practical choice.
However, none of these works has studied graph classification under split data ownership.
Our conference paper~\cite{liu2026privacy} studied this setting for cyberattack detection on the IEEE 34-bus and 123-bus feeders and unlearned the withdrawn charging power.
This paper turns to cyberattack localization and compares FT with second-order GU methods.
To our knowledge, no prior work has used GU to examine what an unlearning request cannot erase when the retained measurements are physically coupled to the withdrawn data.

This paper explores GU for EVCS cyberattack localization with case studies on the IEEE 34-bus, 123-bus, and 8500-node feeders.
The contributions of this work are as follows:
\begin{itemize}
    \item We formulate the withdrawal of EV charging power as a part-feature-level unlearning problem for cyberattack localization, a graph-level multi-label classification task.
    The charging power of the requested EVCSs is zeroed in every training graph, while the voltages and the topology remain.
    We benchmark FT against GIF and IDEA on three distribution feeders with three GNN backbones, Graph Attention Network (GAT)~\cite{velickovic2018gat}, Graph Convolutional Network (GCN)~\cite{kipf2017gcn}, and Graph Isomorphism Network (GIN)~\cite{xu2019gin}.
    \item We show that a withdrawal of charging power leaves a physics-induced leakage floor.
    A GNN retrained without the withdrawn power still localizes cyberattacks at the forgotten EVCSs well above random guessing.
    Power flow analysis explains how this leakage spreads over the feeder topology and decays with the electrical distance from each forgotten EVCS.
    \item We find that GIF and IDEA stay far from retraining, while FT matches it after enough epochs.
    The more EVCSs withdraw, the more epochs FT needs, and on the two smaller feeders a full withdrawal requires running FT to convergence, at a cost close to retraining.
    For unlearning requests that arrive one after another, sequential FT over every request that a feeder can receive loses little, so a DSO can serve them without periodic retraining.
\end{itemize}

The remainder of this paper is organized as follows.
\autoref{sec:methodology} formulates the EVCS cyberattack localization and part-feature-level unlearning problem.
\autoref{sec:experiment} quantifies the physics-induced leakage floor and evaluates the GU methods against FR.
\autoref{sec:conclusion} concludes the paper.

\section{Methodology}
\label{sec:methodology}

\begin{figure*}[thb!]
    \centering
    \includegraphics[width=0.92\textwidth]{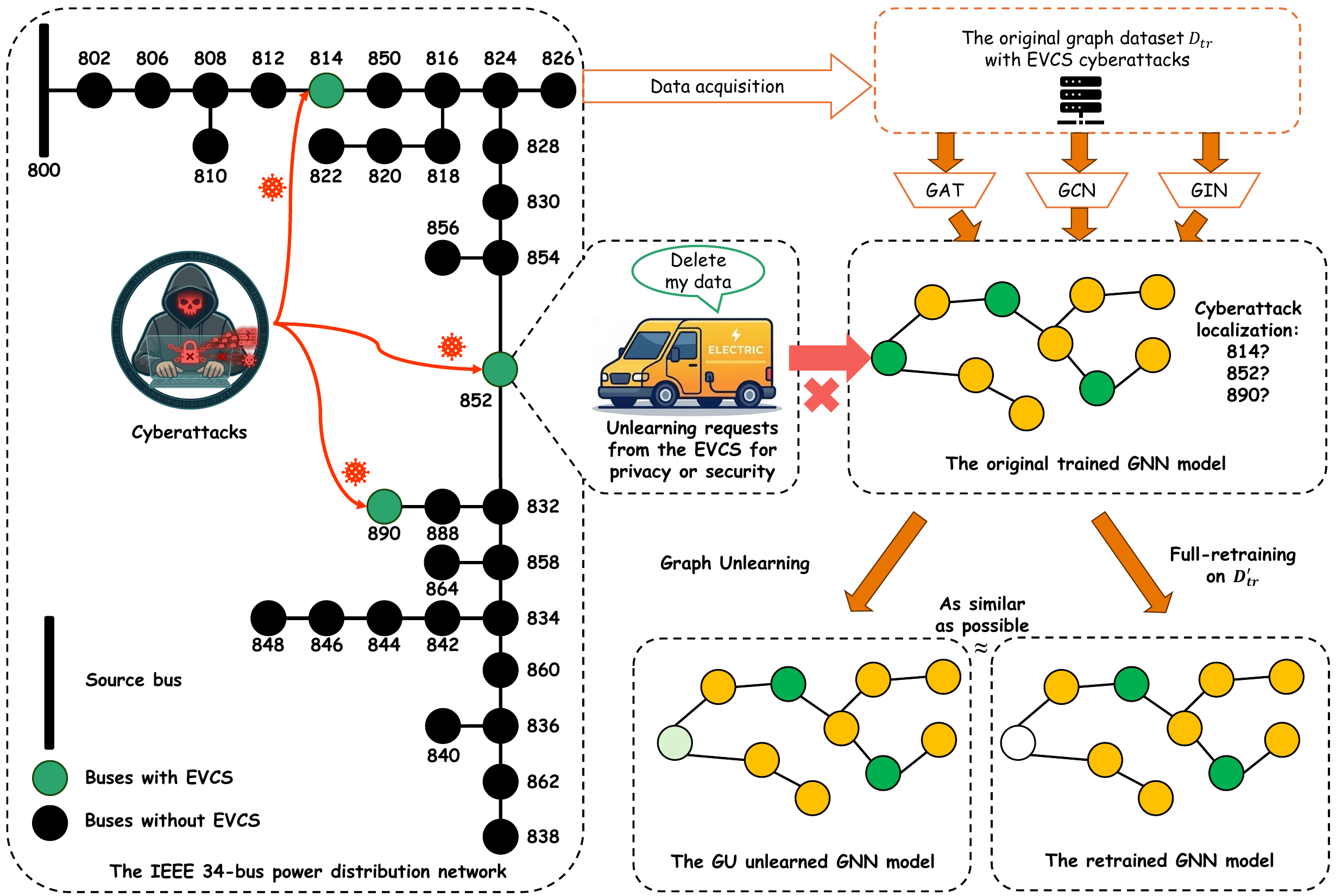}
    \caption{The framework of GU for EVCS cyberattack localization.}
    \label{fig:framework}
\end{figure*}

\subsection{Problem Formulation}

We model each daily feeder snapshot as a static graph $\G = (\V, \E, \mathbf{X})$, where $\V$ is the set of buses with $N = |\V|$, $\E$ is the set of lines or in-line transformers with $M = |\E|$, and $\mathbf{X} \in \R^{N \times F}$ is the node feature matrix.
The node feature of bus $v$ concatenates a voltage block and a charging power block,
\begin{equation}
    \mathbf{x}_v = [\,\mathbf{v}_v \,\|\, \mathbf{p}_v\,] \in \R^{F}
    \label{eq:node_feat}
\end{equation}
where $\mathbf{v}_v$ is the phase-averaged bus voltage and $\mathbf{p}_v$ is the EVCS charging power, both sampled every five minutes over the day.
Each block holds 288 samples, so $F = 576$.
The charging power block $\mathbf{p}_v$ is nonzero only at buses with EVCS, since non-EVCS buses carry no charging measurement.
Voltage and charging power are standardized with separate scalers fit on the training set, with one mean and one standard deviation for each bus and time step.

\textbf{Localization task.}
Cyberattack localization is a graph-level multi-label classification problem.
A GNN $f_\theta : \G \rightarrow \R^{K}$ maps a feeder snapshot to $K$ logits, one per EVCS bus, where $\V_{\text{EVCS}} \subseteq \V$ is the set of EVCS buses and $K = |\V_{\text{EVCS}}|$.
The target $\mathbf{y} \in \{0,1\}^{K}$ is a multi-hot vector with $y_k = 1$ when EVCS $k$ is under attack.
We train $f_\theta$ by minimizing the average loss over a training set $\D_{\text{tr}}$,
\begin{equation}
    \theta^\star = \arg\min_{\theta} \frac{1}{|\D_{\text{tr}}|}
    \sum_{(\G_i, \mathbf{y}_i) \in \D_{\text{tr}}} \ell\big(f_\theta(\G_i), \mathbf{y}_i\big)
    \label{eq:train_obj}
\end{equation}
where $\ell$ is a per-EVCS weighted binary cross-entropy (BCE),
\begin{equation}
    \ell(\hat{\mathbf{z}}, \mathbf{y}) = -\sum_{k=1}^{K}
    \big[\, w_k\, y_k \log \hat{y}_k + (1 - y_k) \log(1 - \hat{y}_k) \,\big]
    \label{eq:bce}
\end{equation}
Here $\hat{y}_k = 1/(1 + e^{-\hat{z}_k})$ is the predicted attack probability of EVCS $k$, and $w_k = n_k^- / n_k^+$ is the negative-to-positive ratio of EVCS $k$ in $\D_{\text{tr}}$, which counters class imbalance.

\textbf{Unlearning requests under split data ownership.}
Control of the data in a distribution feeder is split between two parties.
The DSO controls the voltage data on every bus, and each EVCS operator controls the charging power record at its own EVCS.
The voltages come from the DSO's own feeder measurements.
The charging power records reach the DSO through data-sharing agreements with the operators, for example in managed charging programs.
Each operator can withdraw its records under the terms of its agreement.
The voltages are retained for grid monitoring, so only the charging power is revocable.
An unlearning request names a subset $\V_f \subseteq \V_{\text{EVCS}}$ of EVCS buses whose charging data must be removed.
We realize the request by zeroing the charging power block at the requested buses while leaving voltage and topology intact,
\begin{equation}
    \mathbf{p}_v \leftarrow \mathbf{0}, \quad \forall v \in \V_f
    \label{eq:forget}
\end{equation}
which yields a modified feature matrix $\mathbf{X}'$ and a modified dataset $\D'$.
This is feature-level unlearning that removes only one feature block of each requested node, which we call part-feature-level unlearning.
Each zero is the training mean of its standardized column, so $\D'$ replaces the withdrawn records with a constant, as in the feature unlearning of~\cite{warnecke2021machine}.
Each bus and time step has its own scaler column, so the withdrawn records do not change the scaling of any input that remains.
We study cumulative scenarios $\mathrm{S}_1, \dots, \mathrm{S}_K$, where $\mathrm{S}_j$ forgets the first $j$ EVCS buses, so the forget set grows as more operators withdraw their charging data.
Each scenario starts from the original parameters $\theta^\star$, and \autoref{sec:sequential} also tests the requests in sequence.
\autoref{fig:framework} gives an overview of the whole pipeline.

\textbf{What the unlearning removes.}
Retraining on $\D'$ gives the exact reference, since the retrained model never sees the withdrawn records.
A contractual withdrawal requires removing their influence, so an approximate method such as FT is judged by how close it comes to this reference.
The voltages stay in the model input, because the DSO controls them and needs them for grid monitoring.
A cyberattack at an EVCS changes the power flow and thus the voltages at the attacked bus and at electrically coupled nearby buses.
The retrained reference therefore still carries information about cyberattacks at a forgotten EVCS.
A DSO can assure an operator that its withdrawn data are no longer used in the model.
It cannot assure that the attack history of the EVCS is forgotten, because the voltages still record it.
\autoref{sec:privacy} quantifies this physics-induced cyberattack information leakage floor.

\subsection{GNN Backbones}

We use three GNN backbones, GAT, GCN, and GIN, with a shared graph-classification architecture.
Each backbone stacks three convolution layers, and every layer applies batch normalization, a nonlinearity, and dropout.
\begin{equation}
    \mathbf{h}_v^{(l+1)} = \phi\Big( \mathrm{BN}\big( \mathrm{AGG}^{(l)}
    \big(\{ \mathbf{h}_u^{(l)} : u \in \N(v) \cup \{v\} \}\big) \big) \Big)
    \label{eq:mp}
\end{equation}
where $\N(v)$ is the neighborhood of $v$, $\mathbf{h}_v^{(0)} = \mathbf{x}_v$, and $\mathbf{h}_v^{(l)} \in \R^{H}$ for $l \geq 1$.
The aggregator $\mathrm{AGG}^{(l)}$ and the activation function $\phi$ differ across backbones.
GCN uses the symmetric-normalized sum with ReLU, GAT uses multi-head attention weights with ELU, and GIN uses a summation followed by a multilayer perceptron with ReLU.
We read out a graph embedding by concatenating mean and max pooling over the final node embeddings,
\begin{equation}
    \mathbf{h}_G = \big[\, \mathrm{MEAN}_{v \in \V}\, \mathbf{h}_v^{(L)} \,\|\,
    \mathrm{MAX}_{v \in \V}\, \mathbf{h}_v^{(L)} \,\big] \in \R^{2H}
    \label{eq:readout}
\end{equation}
and score each EVCS from the embedding of its own bus together with this graph embedding,
\begin{equation}
    \hat{z}_k = \mathbf{w}_2^{\top}\, \phi\big(\mathbf{W}_1 \big[\, \mathbf{h}_{v_k}^{(L)} \,\|\, \mathbf{h}_G \,\big]\big) + b_k
    \label{eq:head}
\end{equation}
where $v_k$ is the bus of EVCS $k$, $\mathbf{W}_1$ and $\mathbf{w}_2$ are shared by all EVCSs, and $b_k$ is a bias per EVCS.
Each EVCS is thus scored from its own bus and its neighborhood rather than from an average over all buses, which matters on a large feeder where a few EVCSs sit among thousands of buses.

An attack-type head adds auxiliary supervision to the shared encoder.
The head $g_{\text{aux}}: \R^{2H} \rightarrow \R^{C}$ predicts the graph-level attack type over $C$ classes, covering the clean case and the attack types present in the data, and the model is trained with a joint loss
\begin{equation}
    \mathcal{L} = \mathcal{L}_{\text{loc}} + \gamma\, \mathcal{L}_{\text{type}}
    \label{eq:joint}
\end{equation}
where $\mathcal{L}_{\text{loc}}$ is the weighted BCE of \eqref{eq:bce}, $\mathcal{L}_{\text{type}}$ is a cross-entropy on the attack type, and $\gamma = 0.5$.
We obtain $\theta^\star$ by optimizing \eqref{eq:joint} with Adam under weight decay.

\subsection{Unlearning by Fine-Tuning}
\label{sec:gdgu}
FT starts from the original parameters $\theta^\star$ and continues training on $\D'_{\text{tr}}$, so the model relearns from the retained data instead of estimating the parameter change around $\theta^\star$.
It minimizes the weighted BCE of \eqref{eq:bce} with Adam at the learning rate and weight decay of the original training, freezes the auxiliary head, clips the gradient norm of each minibatch, and keeps the checkpoint with the best validation ROC-AUC on the modified data.
We run FT for $E \in \{10, 25, 50, 100\}$ epochs, written FT~$E$.
These budgets are set in advance, not chosen by the distance to retraining, since the DSO has no retrained model when it unlearns.
FT to convergence has no fixed budget.
It trains from $\theta^\star$ with the procedure of the original training and stops by the same rule once the validation ROC-AUC no longer improves, so it is a warm start of retraining.

\autoref{alg:gdgu} summarizes FT.
FT gives no theoretical bound on its distance to retraining, so \autoref{sec:experiment} measures this distance.

\begin{algorithm}[thb!]
\caption{Unlearning by Fine-Tuning (FT)}
\label{alg:gdgu}
\begin{algorithmic}[1]
\REQUIRE original model $f_{\theta^\star}$; modified training set $\D'_{\text{tr}}$ and validation set $\D'_{\text{val}}$, with the charging power blocks zeroed at $\V_f$ as in \eqref{eq:forget}; epochs $E$; learning rate $\eta$
\ENSURE unlearned model $f_{\theta_{\text{best}}}$
\STATE $\theta_u \leftarrow \theta^\star$; freeze the auxiliary head
\STATE $\theta_{\text{best}} \leftarrow \theta_u$, \quad $r_{\text{best}} \leftarrow 0$
\FOR{$e = 1$ to $E$}
    \FOR{each minibatch $\B \subset \D'_{\text{tr}}$}
        \STATE $\theta_u \leftarrow$ Adam step with rate $\eta$ on the weighted BCE over $\B$, gradient norm clipped to 5
    \ENDFOR
    \STATE $r \leftarrow$ mean ROC-AUC of $f_{\theta_u}$ over the EVCSs on $\D'_{\text{val}}$
    \IF{$r > r_{\text{best}}$}
        \STATE $\theta_{\text{best}} \leftarrow \theta_u$, \quad $r_{\text{best}} \leftarrow r$
    \ENDIF
\ENDFOR
\RETURN $f_{\theta_{\text{best}}}$
\end{algorithmic}
\end{algorithm}

\subsection{Baselines: GIF and IDEA}

We compare FT against two influence-function-based GU baselines.
GIF estimates the parameter change with the inverse Hessian $\mathbf{H}^{-1}$ of the training loss at $\theta^\star$.
Let
\begin{equation}
    \mathbf{g}(\D; \theta) = \frac{1}{|\D|} \sum_{(\G_i, \mathbf{y}_i) \in \D}
    \nabla_\theta\, \ell\big(f_\theta(\G_i), \mathbf{y}_i\big)
    \label{eq:batch_grad}
\end{equation}
denote the average loss gradient over a dataset $\D$.
GIF targets the vector $\mathbf{b} = \mathbf{g}(\D_{\text{tr}}; \theta^\star) - \mathbf{g}(\D'_{\text{tr}}; \theta^\star)$ and approximates $\mathbf{H}^{-1}\mathbf{b}$ with a truncated Neumann series,
\begin{equation}
    \mathbf{q}_0 = \mathbf{b}, \qquad
    \mathbf{q}_j = \mathbf{b} + (1 - \beta)\,\mathbf{q}_{j-1} - \tfrac{1}{s}\,\mathbf{H}\mathbf{q}_{j-1}
    \label{eq:neumann}
\end{equation}
where $\beta$ is a damping term, $s$ is a scaling factor, and each iteration evaluates a Hessian-vector product (HVP).
After $T$ iterations, the parameters are updated by $\theta \leftarrow \theta^\star + \mathbf{q}_T / s$, followed by batch-normalization recalibration.
IDEA, as implemented in OpenGU, adds Gaussian noise of standard deviation $\sigma$ to the updated parameters before the batch-normalization recalibration.
Without the noise, IDEA reduces to GIF, so we set $\sigma$ to the value that its certification bound requires at $\epsilon = 1$ and $\delta = 10^{-4}$, under the default constants of OpenGU.

\begin{figure*}[!t]
    \centering
    \begin{minipage}[t]{0.48\textwidth}
        \centering
        \includegraphics[width=0.95\linewidth]{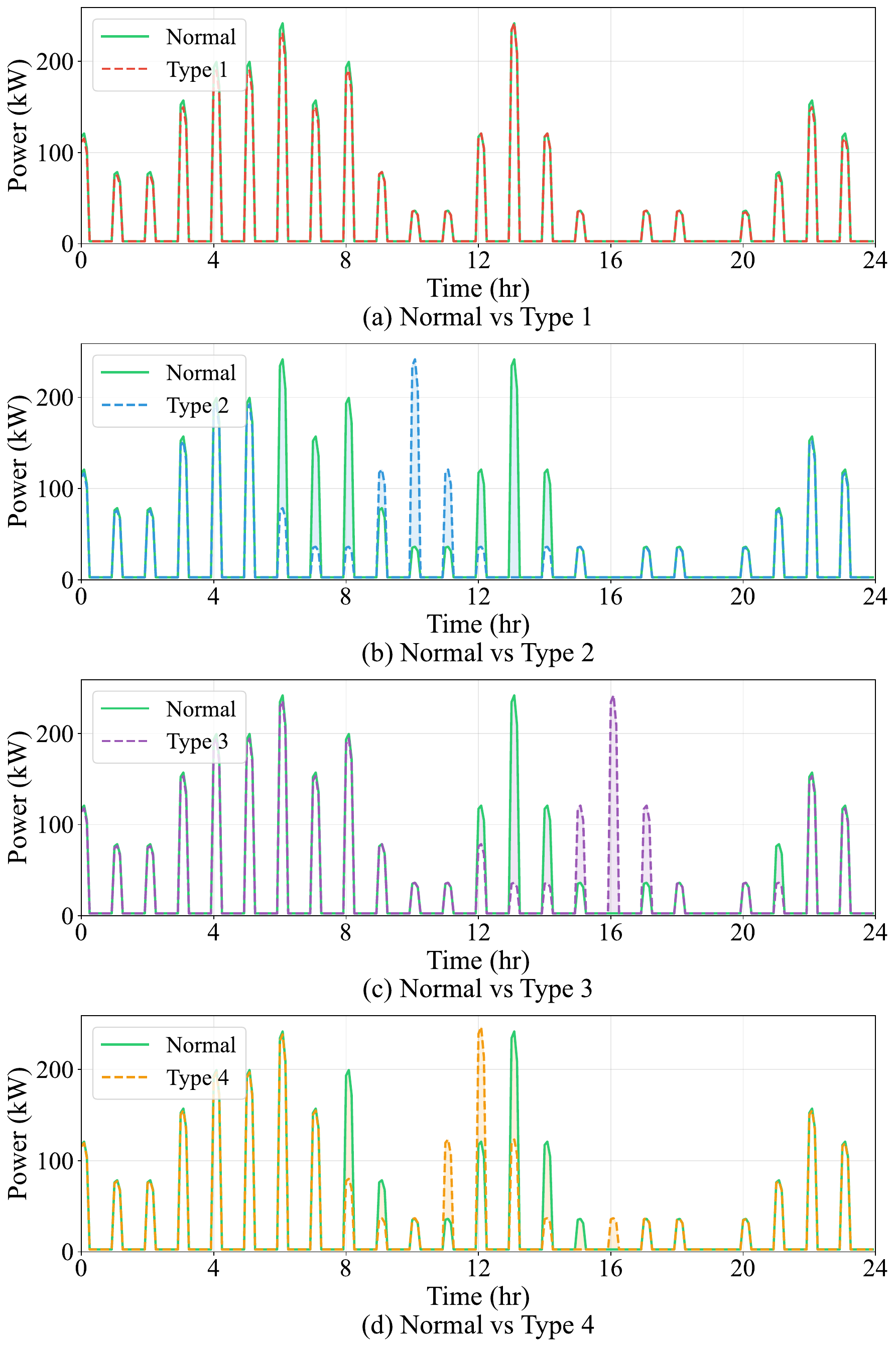}
        \caption{The EVCS charging power of the IEEE 34-bus feeder under normal operation and the four CMA types.}
        \label{fig:attack_power}
    \end{minipage}\hfill
    \begin{minipage}[t]{0.48\textwidth}
        \centering
        \includegraphics[width=0.95\linewidth]{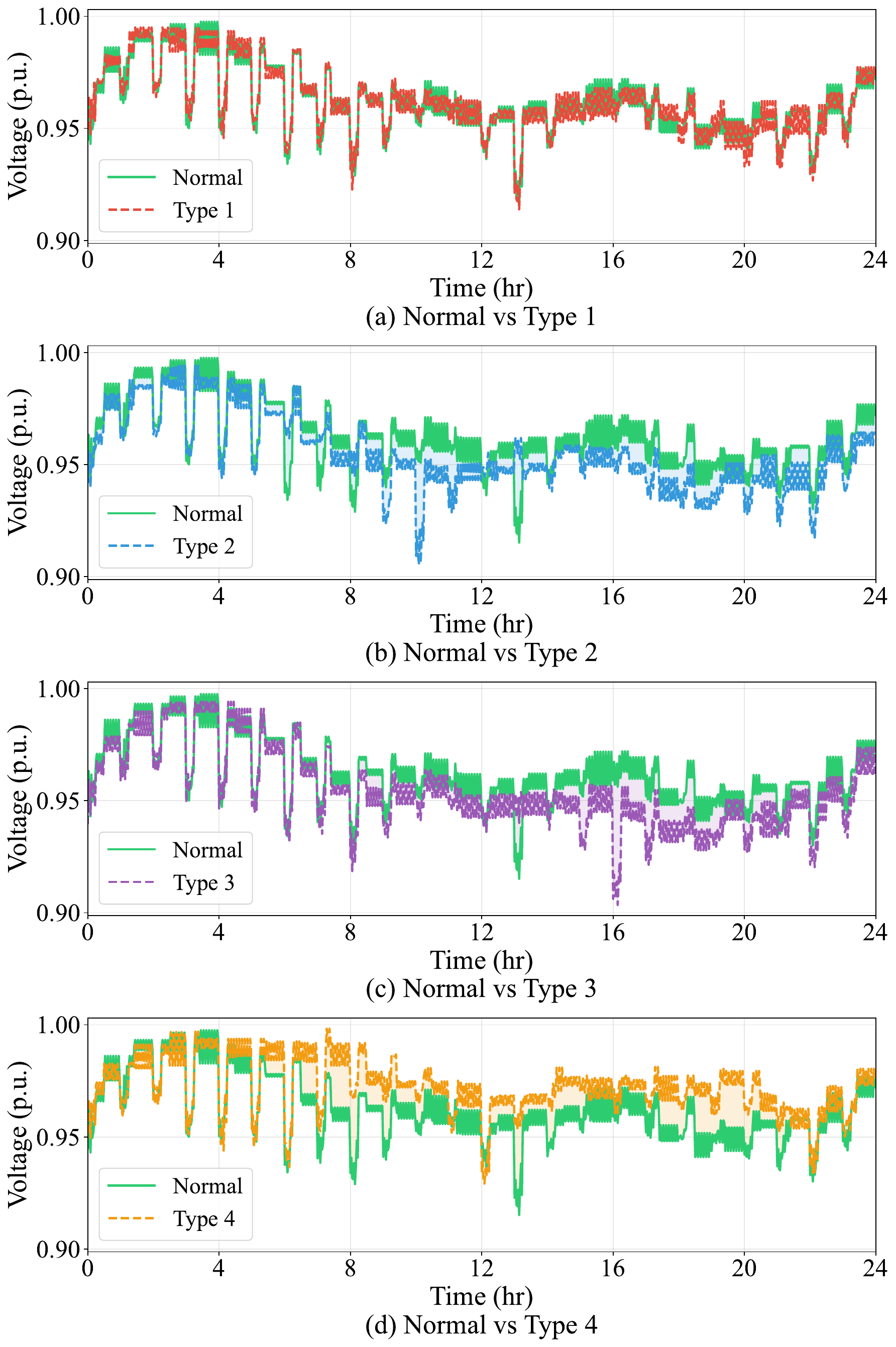}
        \caption{The bus voltage of the IEEE 34-bus feeder under normal operation and the four CMA types.}
        \label{fig:attack_voltage}
    \end{minipage}
\end{figure*}

\section{Experimental Results}
\label{sec:experiment}

\subsection{Experimental Setup}

\textbf{Datasets and scenarios.}
PowerBench~\cite{uddin2026powerbench} is an open benchmark suite for monitoring and security in power distribution networks.
It covers datasets across tasks such as outage detection, cyberattack localization, and state estimation, built on high-fidelity simulations of standard IEEE test feeders.
This work focuses on its EVCS cyberattack dataset, which records daily snapshots of the IEEE 34-bus, 123-bus, and 8500-node feeders under normal operation and four CMA types, with EVCSs at known buses and ground-truth labels for every snapshot.
PowerBench derives the charging profiles of this dataset from real-world EVCS usage data in Texas.
The four CMA types manipulate the charging power as follows.
\begin{itemize}
    \item Type~1 deviates only slightly from the normal profile and is the hardest to detect.
    \item Type~2 shifts the charging power demand to an earlier time of day.
    \item Type~3 shifts the charging power demand to a later time of day.
    \item Type~4 reshapes the daily peak by injecting a surge that exceeds the normal maximum.
\end{itemize}
\autoref{fig:attack_power} shows the four CMA types on the IEEE 34-bus feeder for the charging power and \autoref{fig:attack_voltage} for the bus voltage.
Every manipulation perturbs the charging power of the targeted EVCS and, through the power flow, the voltages of its bus and of electrically coupled buses.
The auxiliary attack-type head therefore predicts $C = 5$ classes on this dataset, the non-attack case and the four CMA types.
\autoref{tab:dataset} summarizes the three test feeders.
They span two orders of magnitude in size, from $37$ to $4876$ buses, and from $3$ to $7$ EVCSs.
We run all scenarios $\mathrm{S}_1$ to $\mathrm{S}_K$ on the 34-bus and 123-bus feeders, and $\mathrm{S}_1$ and $\mathrm{S}_7$ on the 8500-node feeder.

\begin{table}[thb!]
\centering
\caption{Properties of the three IEEE test feeders in the PowerBench EVCS dataset.}
\label{tab:dataset}
\footnotesize
\setlength{\tabcolsep}{4pt}
\begin{tabular*}{0.94\columnwidth}{@{\extracolsep{\fill}}lccc@{}}
\toprule
Property & 34-bus & 123-bus & 8500-node \\
\midrule
Buses / nodes $N$              & 37   & 132  & 4876 \\
Lines / edges $M$           & 36   & 131  & 4874 \\
Node features $F$   & 576   & 576   & 576 \\
Graphs                 & 2000 & 4000 & 1000 \\
EVCS buses $K$         & 3    & 5    & 7 \\
Unlearning scenarios   & S1--S3 & S1--S5 & S1, S7 \\
GNN backbones              & GAT/GCN/GIN & GAT/GCN/GIN & GCN/GIN \\
\bottomrule
\end{tabular*}
\end{table}

\textbf{Baselines.}
We use GAT, GCN, and GIN on the 34-bus and 123-bus feeders, and GCN and GIN on the 8500-node feeder.
A trial fixes the feeder, scenario, backbone, and seed, and runs every method on the same split.
These methods are FT~10, 25, 50, and 100, FT to convergence, GIF, and IDEA.
The references are the original model and FR on $\D'_{\text{tr}}$.
We score the original model with the withdrawn power as before the request, written Original (with power), and on $\D'$ without any update, written Original (power removed).
Since every method starts from $\theta^\star$, Original (power removed) is where each method stands before its first update, and FR is the model it should match.

\textbf{Hyperparameters.}
We split the input data into $70\%/15\%/15\%$ for training, validation, and testing, stratified on the multi-label combination.
Every configuration runs with 10 random seeds.
All GNNs use Adam with a learning rate and weight decay of $10^{-4}$, minibatches of 64 graphs, a hidden dimension $H = 128$, three convolution layers, and dropout $0.3$.
Training stops once the validation ROC-AUC has not improved for 50 epochs.
GIF and IDEA use $T = 50$ Neumann iterations with damping $\beta = 0.05$ and scaling factor $s = 200$.

The 34-bus and 123-bus experiments ran on a workstation with RTX 4090 GPUs, and the 8500-node experiments ran on an H100 GPU from high-performance computing resources~\cite{utdhpc}.
All models are implemented in PyTorch Geometric, and the code is available at {\fontencoding{T1}\selectfont\url{https://github.com/lnhfrank/GU_EV_localization}}.

\textbf{Metrics.}
Each EVCS is scored by its ROC-AUC and its F1 score on the test graphs, where every model except Original (with power) sees the withdrawn power zeroed.
For the F1 score, an EVCS counts as attacked when its predicted probability exceeds $0.5$.
The forgotten-station ROC-AUC and F1 score average these scores over the forgotten EVCSs.
For FR, the forgotten-station ROC-AUC is the leakage floor.
An unlearning method should come close to FR in both, not fall toward random guessing.
Within each scenario, we pair the trials of a method with those of FR by backbone and seed, and a two-sided Wilcoxon signed-rank test~\cite{wilcoxon1992individual} checks whether its F1 score differs from that of FR at the 0.05 level.
A loss-based membership inference attack (MIA)~\cite{wu2021adapting} tells the training graphs from the test graphs by the loss at the forgotten EVCSs, with the withdrawn power zeroed in both.
Since the training graphs stay in $\D'_{\text{tr}}$, the attack cannot show whether the withdrawn power was used, so its AUC is only compared with that of FR, next to the two scores above, and does not decide whether a method comes close to FR.

\begin{table*}[t]
\centering
\caption{The forgotten-station F1 score.}
\label{tab:agreement}
\footnotesize
\setlength{\tabcolsep}{5pt}
\begin{tabular*}{0.94\textwidth}{@{\extracolsep{\fill}}lcccccc@{}}
\toprule
\multirow{2}{*}[-0.6ex]{Method} & \multicolumn{2}{c}{IEEE 34-bus} & \multicolumn{2}{c}{IEEE 123-bus} & \multicolumn{2}{c}{IEEE 8500-node} \\
\cmidrule(lr){2-3} \cmidrule(lr){4-5} \cmidrule(lr){6-7}
 & S1 (single) & S3 (all) & S1 (single) & S5 (all) & S1 (single) & S7 (all) \\
\midrule
Original (with power) & 0.985$\pm$0.009 & 0.982$\pm$0.009 & 0.992$\pm$0.006 & 0.988$\pm$0.007 & 0.981$\pm$0.013 & 0.981$\pm$0.009 \\
Original (power removed) & 0.681$\pm$0.009 & 0.687$\pm$0.006 & 0.606$\pm$0.030 & 0.637$\pm$0.006 & 0.539$\pm$0.045 & 0.574$\pm$0.019 \\
\midrule
GIF & 0.588$\pm$0.061 & 0.333$\pm$0.156 & 0.540$\pm$0.076 & 0.198$\pm$0.151 & 0.464$\pm$0.068 & 0.018$\pm$0.038 \\
IDEA & 0.379$\pm$0.298 & 0.382$\pm$0.303 & 0.364$\pm$0.277 & 0.360$\pm$0.283 & 0.369$\pm$0.270 & 0.377$\pm$0.263 \\
FT 10 & 0.682$\pm$0.048 & 0.721$\pm$0.026 & 0.689$\pm$0.038 & 0.655$\pm$0.016 & 0.533$\pm$0.065 & 0.571$\pm$0.033 \\
FT 25 & 0.749$\pm$0.045 & 0.747$\pm$0.023 & 0.716$\pm$0.030 & 0.678$\pm$0.015 & 0.552$\pm$0.055 & \textbf{0.590$\pm$0.031} \\
FT 50 & 0.804$\pm$0.035 & 0.781$\pm$0.021 & 0.749$\pm$0.022 & 0.705$\pm$0.016 & 0.553$\pm$0.055 & 0.602$\pm$0.025 \\
FT 100 & \textbf{0.831$\pm$0.030} & 0.807$\pm$0.020 & 0.780$\pm$0.018 & 0.739$\pm$0.011 & 0.548$\pm$0.061 & 0.607$\pm$0.027 \\
FT to convergence & 0.838$\pm$0.028 & \textbf{0.829$\pm$0.022} & \textbf{0.792$\pm$0.030} & \textbf{0.789$\pm$0.013} & \textbf{0.544$\pm$0.061} & 0.604$\pm$0.030 \\
\midrule
FR & 0.832$\pm$0.031 & 0.826$\pm$0.024 & 0.795$\pm$0.032 & 0.782$\pm$0.014 & 0.544$\pm$0.059 & 0.586$\pm$0.037 \\
\bottomrule
\end{tabular*}
\par\smallskip\parbox{0.94\textwidth}{\footnotesize \textbf{Bold} marks the GU method closest to FR. All the numbers are mean$\pm$sd values over the GNN backbones and 10 seeds.}
\end{table*}

\subsection{Physics-Induced Leakage Floor}
\label{sec:privacy}

\begin{figure}[!t]
    \centering
    \includegraphics[scale=0.49]{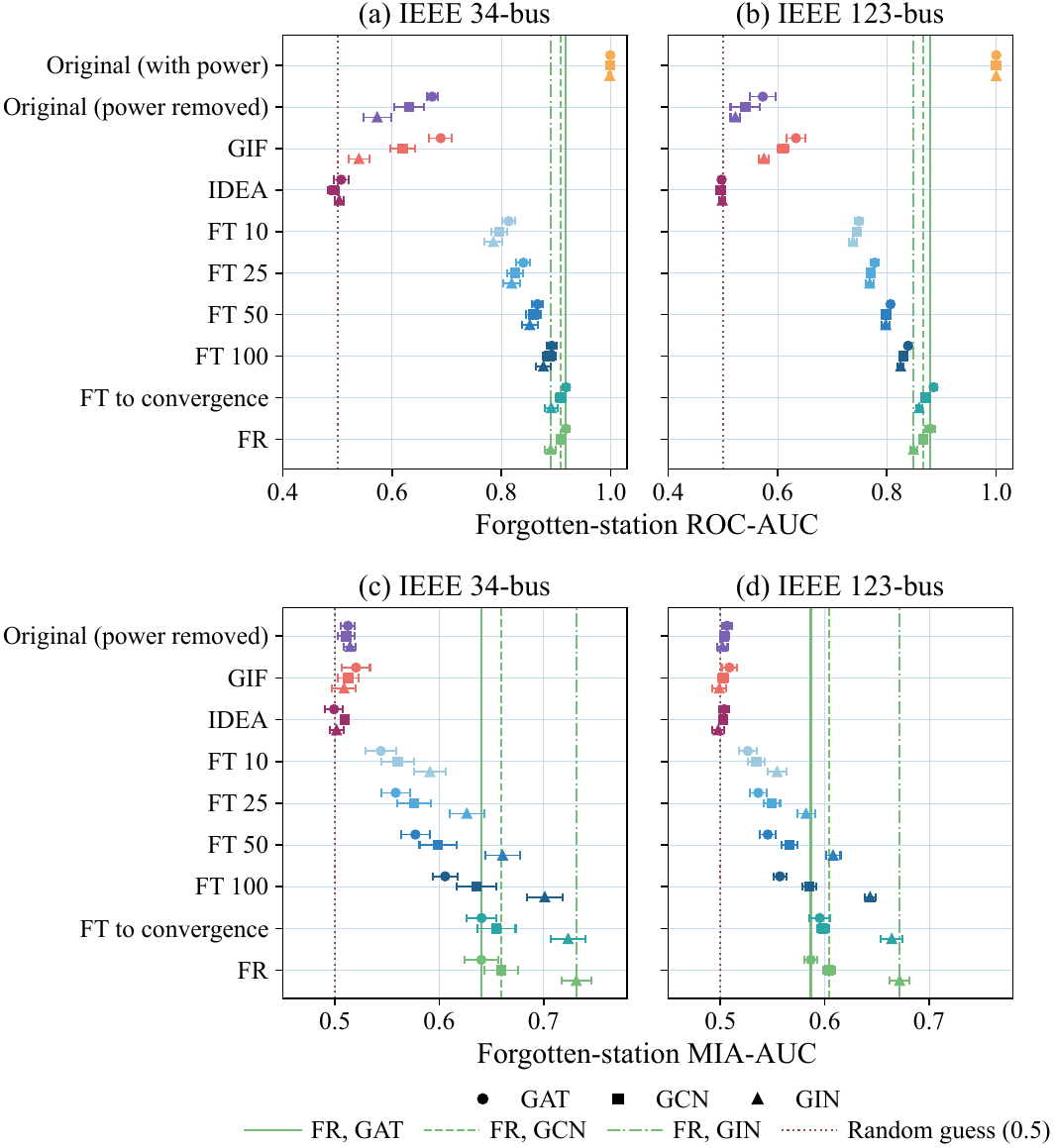}
    \caption{The forgotten-station ROC-AUC and MIA-AUC of every method when every EVCS withdraws its charging power data.}
    \label{fig:floor_methods}
\end{figure}

Retraining without the withdrawn charging power leaves a high physics-induced cyberattack information leakage floor.
When every EVCS withdraws its data, FR still localizes cyberattacks at the forgotten EVCSs with a ROC-AUC of 0.906 on the 34-bus feeder and 0.865 on the 123-bus feeder, as \autoref{fig:floor_methods} shows.
The original model scores 0.999 and 0.998 with the charging power and drops to 0.626 and 0.546 once the withdrawn power is removed, because it reads the attacks from the charging power.
FR instead learns them from the voltages.
Across the scenarios the floor ranges from 0.906 to 0.920 on the 34-bus feeder and from 0.864 to 0.896 on the 123-bus feeder.
On the 8500-node feeder it is lower, 0.616 in $\mathrm{S}_1$ and 0.667 in $\mathrm{S}_7$, with standard deviations of 0.041 and 0.035, but still above random guessing.

\begin{figure}[!t]
    \centering
    \includegraphics[scale=0.49]{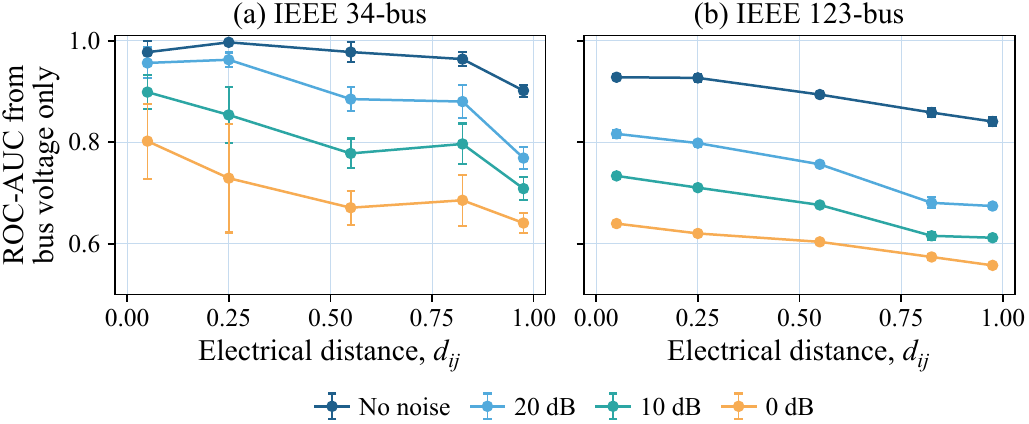}
    \caption{The voltage-only ROC-AUC against the electrical distance from the EVCS.}
    \label{fig:floor_distance}
\end{figure}

\begin{figure*}[!t]
    \centering
    \includegraphics{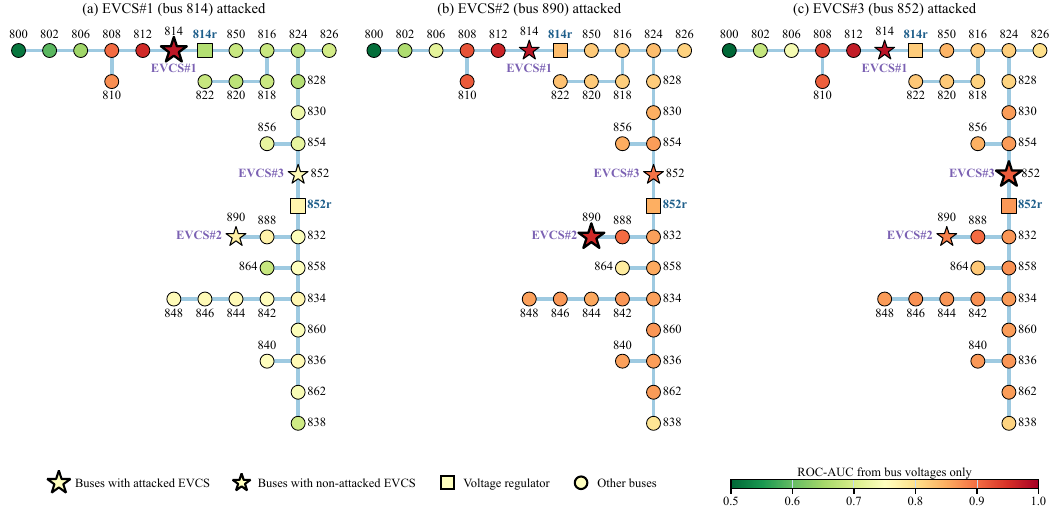}
    \caption{The voltage-only ROC-AUC of each bus on the IEEE 34-bus feeder.}
    \label{fig:topology}
\end{figure*}

The linearized distribution flow (LinDistFlow) model of a radial feeder~\cite{wu1989network} explains where this information comes from.
A change $\Delta P_j$ of the charging power at EVCS $j$ changes the voltage at bus $i$ by
\begin{equation}
    \Delta V_i \approx -R_{ij}\, \Delta P_j
    \label{eq:lindistflow}
\end{equation}
where $R_{ij}$ is the resistance, in per unit, of the path that buses $i$ and $j$ share toward the substation.
A voltage regulator holds its output voltage, so $R_{ij}$ counts only the shared path downstream of the last regulator on it, and $R_{ij} = 0$ for a bus behind a regulator that the path of EVCS $j$ does not pass.
We define the electrical distance from EVCS $j$ as
\begin{equation}
    d_{ij} = \max\left(0,\, 1 - \frac{R_{ij}}{R_{jj}}\right)
    \label{eq:distance}
\end{equation}
It is 0 at the EVCS and at the buses downstream of it up to the next regulator, and 1 at a bus behind a regulator that the path of the EVCS does not pass.
The maximum sets $d_{ij}$ to 0 for buses upstream of the regulator in front of an EVCS, which respond more strongly than the EVCS itself.
On the bus and EVCS pairs that no regulator separates, the voltage response to the charging power in the data follows \eqref{eq:lindistflow} with a correlation of 0.96 on the 34-bus feeder and 0.89 on the 123-bus feeder, against 0.36 and 0.45 when the regulators are ignored.
The buses behind a regulator still show 65\% and 42\% of the mean response of the other buses, so a regulator blocks only part of the signal.

\autoref{fig:floor_distance} shows how this information decays with the electrical distance.
For every bus and EVCS, a logistic regression on the voltage of that single bus predicts whether the EVCS is attacked.
We call its ROC-AUC under 5-fold cross-validation the voltage-only ROC-AUC.
Without noise, the voltage-only ROC-AUC falls from 0.978 for the buses with $d_{ij} \le 0.1$ to 0.904 for those with $d_{ij} \ge 0.9$ on the 34-bus feeder, and from 0.928 to 0.841 on the 123-bus feeder.
We also add white Gaussian noise to the bus voltages at a signal-to-noise ratio (SNR) of 20, 10, and 0~dB, taken against the median standard deviation of the bus voltages.
With noise, the voltage-only ROC-AUC falls faster with the distance.
At an SNR of 10~dB the two groups reach 0.899 and 0.711 on the 34-bus feeder and 0.734 and 0.612 on the 123-bus feeder.
Since FR uses the voltages of all buses, the floor stays well above random guessing.
With every EVCS withdrawn, it is 0.873 and 0.806 at 10~dB and 0.795 and 0.733 at 0~dB.
A logistic regression on the voltages of all buses matches or exceeds the floor at every noise level, for example 0.971 against 0.906 and 0.979 against 0.865 without noise, and 0.732 against 0.733 on the 123-bus feeder at 0~dB.
The floor is therefore a property of the voltages, not of the GNN that measures it.

\begin{figure}[!t]
    \centering
    \includegraphics{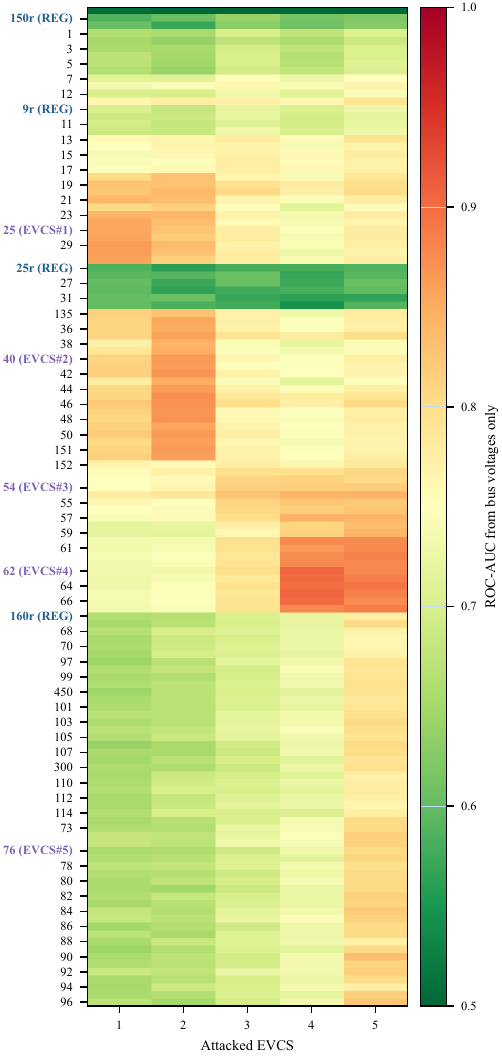}
    \caption{The voltage-only ROC-AUC of each bus on the IEEE 123-bus feeder.}
    \label{fig:heatmap}
\end{figure}

We map the voltage-only ROC-AUC at 20~dB on the 34-bus feeder in \autoref{fig:topology}, with one panel per attacked EVCS, and list it for every bus of the 123-bus feeder in \autoref{fig:heatmap}, with one column per attacked EVCS and the buses in feeder order from the substation.
On both feeders, each attacked EVCS is most visible on the buses around it, and the buses next to the substation, which share little line resistance with any EVCS, reveal the least.
On the 123-bus feeder, the buses behind a regulator that the path of the EVCS does not pass reveal less than the others, but still more than random guessing on average.
On the 34-bus feeder, the buses behind the regulators 814r and 852r still reveal the attacks.

\begin{figure}[!t]
    \centering
    \includegraphics[scale=0.4888]{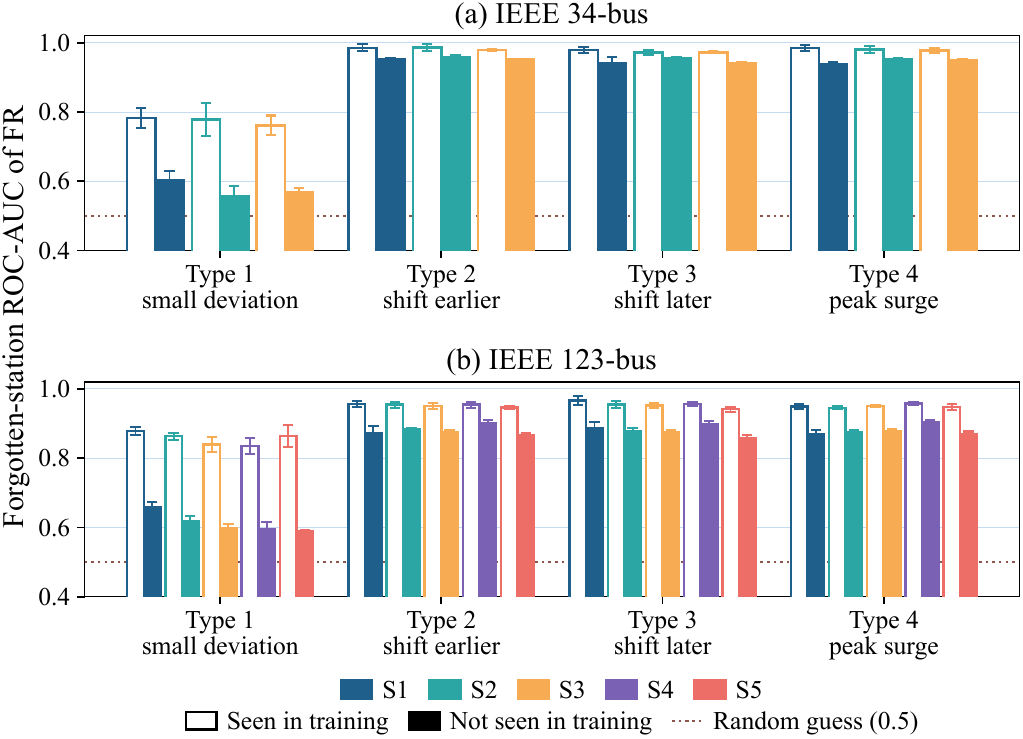}
    \caption{The forgotten-station ROC-AUC of FR for each attack type, seen or not seen in training.}
    \label{fig:holdout}
\end{figure}

FR localizes the cyberattacks at the forgotten EVCSs even for an attack type it has not seen in training.
In \autoref{fig:holdout} we train FR with GCN without one attack type and test it on that type.
For the time shifts and the peak surge, Types~2 to~4, its forgotten-station ROC-AUC is only 0.018 to 0.083 lower than when the type is seen, and stays at 0.858 or above on both feeders.
For Type~1, the small deviation, it is 0.181 to 0.275 lower but stays above random guessing, at 0.558 or above.
Its small change of the charging power leaves only a small change in the voltages, so the model cannot read the attack from the voltages as well as it can for the other types.

For a single request in $\mathrm{S}_1$, the operator of the withdrawn EVCS faces a floor of 0.914 on the 34-bus feeder and 0.896 on the 123-bus feeder.
A DSO that retrains can assure this operator that the model no longer depends on its charging power records, but not that the cyberattacks at its EVCS stay hidden, because the voltages that the DSO keeps reveal them.

\subsection{Unlearning Methods against Retraining}
\label{sec:agreement}

The GU methods are GIF, IDEA, and FT with 10 to 100 epochs or run to convergence.
Each starts from Original (power removed) and is judged by how close its forgotten-station ROC-AUC and F1 score come to those of FR.
\autoref{tab:agreement} gives the forgotten-station F1 score of each method.
It averages over the seeds and the backbones, since the leakage floor and the ranking of the methods hold for each backbone.

GIF barely improves the original model.
With one EVCS withdrawn on the 34-bus feeder, its forgotten-station ROC-AUC is 0.611, close to 0.601 for Original (power removed) and far below the floor of 0.914, and its F1 score is 0.588 against 0.832 for FR.
\autoref{fig:floor_methods} shows the same gap with every EVCS withdrawn.
GIF stays well below FR on every feeder and in both scenarios, because an update around $\theta^\star$ cannot relearn the cyberattacks from the voltages.
With every EVCS withdrawn, it counts a forgotten EVCS as attacked in only 0.188, 0.100, and 0.007 of the test graphs on the three feeders, while FR does so in about half of them, so its F1 score falls to 0.333, 0.198, and 0.018.
Original (power removed) does the opposite and counts it as attacked in 0.936 and 0.885 of the test graphs in $\mathrm{S}_1$ on the 34-bus and 123-bus feeders, so it keeps an F1 score of 0.681 and 0.606, although its ROC-AUC is only 0.601 and 0.504.

In our experiments, a score near random guessing marks a broken model rather than strong forgetting.
IDEA reaches a forgotten-station ROC-AUC of 0.499 on the 34-bus feeder in $\mathrm{S}_1$, which would count as complete forgetting if judged against random guessing.
Its noise, however, also brings the ROC-AUC on the retained EVCSs down to 0.500, against 0.998 for FR, so the model no longer localizes any cyberattack, and its F1 score drops to 0.379 against 0.832 for FR.
The MIA misleads in the same way, as \autoref{fig:floor_methods}(c) and (d) show.
With every EVCS withdrawn, its AUC is 0.501 to 0.514 for Original (power removed), GIF, and IDEA, close to random guessing, whereas FR reaches 0.677 and 0.621 on the 34-bus and 123-bus feeders, and FT approaches these values with more epochs.

FT relearns the attacks, but only with enough epochs.
On the 34-bus feeder in $\mathrm{S}_1$, its forgotten-station ROC-AUC rises from 0.791 after 10 epochs to 0.866 after 25 and 0.920 after 100, against 0.914 for FR.
A score below FR looks like stronger forgetting, but it means that FT has not yet relearned the attacks from the voltages.
With one EVCS withdrawn, FT~100 matches FR on the 34-bus feeder, with an F1 score of 0.831 against 0.832.
On the 123-bus feeder it stays slightly below FR, 0.780 against 0.795, a gap that the paired test finds significant.
After a full withdrawal, FT~100 stays below FR on these feeders, 0.807 against 0.826 and 0.739 against 0.782.
Only FT to convergence reaches FR, with 0.829 and 0.789, and ends slightly but significantly above it on the 123-bus feeder.
The epochs FT needs therefore grow with the size of the withdrawal.

On the retained EVCSs, every method except IDEA stays within 0.002 of the ROC-AUC of FR in $\mathrm{S}_1$ on every feeder, so one withdrawal does not hurt the localization at the other EVCSs.
On the 8500-node feeder, the floor is low, 0.616 in $\mathrm{S}_1$, and the F1 score, 0.539 against 0.544, no longer separates Original (power removed) from FR.
In ROC-AUC, FT~25 already reaches FR in $\mathrm{S}_1$, 0.627 against 0.616, whereas Original (power removed) and GIF stay at 0.517 and 0.529, so the ranking of the methods holds on a feeder of 4876 buses.
In $\mathrm{S}_7$, FT~50, FT~100, and FT to convergence end slightly above FR, with F1 scores of 0.602 to 0.607 against 0.586.
Since FT to convergence follows the procedure of FR and differs from it only in starting from $\theta^\star$, its excess over FR is a residue of the original model, which learned these EVCSs with their charging power.

\subsection{Sequential Unlearning Requests}
\label{sec:sequential}

The scenarios above start every request from $\theta^\star$.
Withdrawals, however, arrive one after another, so we also run them in sequence.
The request of $\mathrm{S}_j$ starts from the model that FT~25 returned for $\mathrm{S}_{j-1}$.
Every step after $\mathrm{S}_1$ ends closer to FR in the forgotten-station ROC-AUC than the same request started from $\theta^\star$.
The gap to FR is $-0.041$ and $-0.037$ at $\mathrm{S}_2$ and $\mathrm{S}_3$ on the 34-bus feeder and $-0.015$ to $-0.038$ at $\mathrm{S}_2$ to $\mathrm{S}_5$ on the 123-bus feeder, whereas the same requests started from $\theta^\star$ give $-0.044$ to $-0.089$.
The retained EVCSs stay within 0.0004 of FR at every step.

\begin{figure}[!t]
    \centering
    \includegraphics[scale=0.4888]{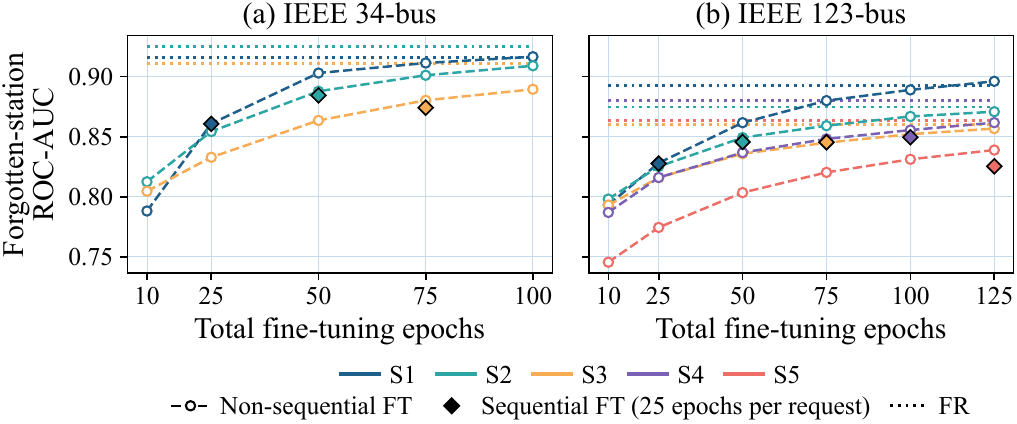}
    \caption{The forgotten-station ROC-AUC against the total fine-tuning epochs of sequential and non-sequential FT.}
    \label{fig:sequential}
\end{figure}

Sequential FT stays closer to FR because the epochs add up along the sequence.
\autoref{fig:sequential} plots each step against the total epochs of FT.
At the same total number of epochs, sequential FT~25 lags slightly behind one FT run from $\theta^\star$, most at $\mathrm{S}_5$ on the 123-bus feeder, $-0.038$ against $-0.025$.
Running the requests in sequence thus costs a little efficiency, a lag of at most 0.013 by the time every EVCS of the feeder has withdrawn.
In our setup, a request withdraws all the charging power of one EVCS, so a feeder receives at most one request per EVCS, and sequential FT needs no periodic retraining to remove a drift.

\subsection{Cost per Unlearning Request}
\label{sec:cost}

\begin{table}[t]
\centering
\caption{The time consumption and peak GPU memory usage per request in $\mathrm{S}_1$ for GCN.}
\label{tab:cost}
\footnotesize
\setlength{\tabcolsep}{3pt}
\begin{tabular*}{0.94\columnwidth}{@{\extracolsep{\fill}}lrrrrrr@{}}
\toprule
\multirow{2}{*}[-0.6ex]{Method} & \multicolumn{2}{c}{34-bus} & \multicolumn{2}{c}{123-bus} & \multicolumn{2}{c}{8500-node} \\
\cmidrule(lr){2-3} \cmidrule(lr){4-5} \cmidrule(lr){6-7}
 & s & MB & s & MB & s & MB \\
\midrule
GIF & 9.8 & 61 & 25.8 & 155 & 130.8 & 4896 \\
IDEA & 9.9 & 61 & 25.7 & 155 & 129.8 & 4896 \\
FT 10 & 1.8 & 43 & 4.1 & 100 & 27.2 & 2985 \\
FT 25 & 4.6 & 43 & 10.4 & 100 & 67.1 & 2985 \\
FT 50 & 8.8 & 43 & 20.5 & 100 & 132.4 & 2985 \\
FT 100 & 15.4 & 43 & 41.4 & 100 & 255.1 & 2985 \\
FT to convergence & 46.9 & 44 & 125.0 & 101 & 320.4 & 2985 \\
FR & 53.6 & 44 & 183.2 & 101 & 583.0 & 2984 \\
\bottomrule
\end{tabular*}
\end{table}

\autoref{tab:cost} gives the median wall-clock time and the peak GPU memory of one request in $\mathrm{S}_1$ for GCN.
The time covers every step that a method needs to return its model, but not the evaluation on the test set.
The time of FT grows in proportion to its epochs, from 4.6~s for FT~25 to 15.4~s for FT~100 on the 34-bus feeder, so its speed advantage over FR, which takes 53.6~s, is the ratio of their epochs rather than a property of the method.
A fair comparison is the time FT needs to reach FR.
For one withdrawn EVCS, FT~100 comes close to FR in 15.4 and 41.4~s on the 34-bus and 123-bus feeders, which is 29\% and 23\% of the time of FR.
A full withdrawal needs FT to convergence, whose median epochs are 79\% and 95\% of those of FR, so it saves little time.
GIF and IDEA take 9.8 to 25.8~s on these feeders, but \autoref{sec:agreement} shows that they stay far from FR.

FT keeps the peak GPU memory of standard training, 43, 100, and 2985~MB on the three feeders, while GIF and IDEA need 61, 155, and 4896~MB, 1.4 to 1.6 times as much, since each Hessian-vector product differentiates the gradient of a minibatch once more.

\FloatBarrier

\section{Conclusion} \label{sec:conclusion}
We study graph unlearning for EVCS cyberattack localization under split data ownership, in which an unlearning request removes only the charging power features of a bus, while voltages and the graph topology are preserved.
Since the voltage data are never revoked, such a request leaves a physics-induced leakage floor.
A model can forget the withdrawn power, but not its physical consequences in the voltages.
Therefore, a model retrained without the withdrawn power still localizes cyberattacks at the forgotten EVCSs well above random guessing, even under measurement noise.
A linearized power flow explains this floor through the line resistance that a bus shares with the EVCS, and a logistic regression on all bus voltages matches or exceeds the GNN, so the floor is a property of the voltages.
Therefore, a DSO can promise an operator that a retrained model no longer uses the withdrawn records, but the voltage telemetry still reveals the attack history of the EVCS.

This study also has limitations.
FT gives no theoretical bound on its distance to retraining, so this distance has to be measured.
FT also keeps a residue of the original model and assumes that the retained data stay available.

Potential future work includes extending the analysis to edge or graph level unlearning requests and to charging data at other EV penetration levels.

\section*{Acknowledgment}
The authors acknowledge the High Performance Computing at The University of Texas at Dallas (HPC@UTD) for providing the GPU computing resources and support.

\bibliographystyle{IEEEtran}
\bibliography{references}

@article{markovic2023machine,
  title={Machine learning for modern power distribution systems: Progress and perspectives},
  author={Markovi{\'c}, Marija and Bossart, Matthew and Hodge, Bri-Mathias},
  journal={Journal of Renewable and Sustainable Energy},
  volume={15},
  number={3},
  pages={032301},
  year={2023},
  publisher={AIP Publishing LLC}
}

@article{liao2021review,
  title={A review of graph neural networks and their applications in power systems},
  author={Liao, Wenlong and Bak-Jensen, Birgitte and Pillai, Jayakrishnan Radhakrishna and Wang, Yuelong and Wang, Yusen},
  journal={Journal of Modern Power Systems and Clean Energy},
  volume={10},
  number={2},
  pages={345--360},
  year={2021},
  publisher={SGEPRI}
}

@article{shaik2024exploring,
  title={Exploring the landscape of machine unlearning: A comprehensive survey and taxonomy},
  author={Shaik, Thanveer and Tao, Xiaohui and Xie, Haoran and Li, Lin and Zhu, Xiaofeng and Li, Qing},
  journal={IEEE Transactions on Neural Networks and Learning Systems},
  volume={36},
  number={7},
  pages={11676--11696},
  year={2024},
  publisher={IEEE}
}

@article{uddin2026powerbench,
title={PowerBench: A Benchmark Suite for Monitoring and Security in Distribution Systems},
author={Md Joshem Uddin and Roshni Anna Jacob and Damilola R. Olojede and Nanhong Liu and Baris Coskunuzer and Jie Zhang},
journal={Journal of Data-centric Machine Learning Research},
year={2026},
}

@inproceedings{jacob2025cyber,
  author    = {Jacob, Roshni Anna and Uddin, Md Joshem and Wang, Jingbo and Coskunuzer, Baris and Zhang, Jie},
  title     = {Cyber Attack Detection in Electric Vehicle Charging Stations Using Topological Data Aided Learning},
  booktitle = {2025 IEEE Power \& Energy Society General Meeting (PESGM)},
  year      = {2025},
  doi       = {10.1109/PESGM52009.2025.11225416},
}

@article{jahangir2024charge,
  title={Charge manipulation attacks against smart electric vehicle charging stations and deep learning-based detection mechanisms},
  author={Jahangir, Hamidreza and Lakshminarayana, Subhash and Poor, H Vincent},
  journal={IEEE Transactions on Smart Grid},
  volume={15},
  number={5},
  pages={5182--5194},
  year={2024},
  publisher={IEEE}
}

@inproceedings{kipf2017gcn,
  author    = {Kipf, Thomas N. and Welling, Max},
  title     = {Semi-Supervised Classification with Graph Convolutional Networks},
  booktitle = {International Conference on Learning Representations (ICLR)},
  year      = {2017},
}

@inproceedings{velickovic2018gat,
  author    = {Veli{\v{c}}kovi{\'c}, Petar and Cucurull, Guillem and Casanova, Arantxa and Romero, Adriana and Li{\`o}, Pietro and Bengio, Yoshua},
  title     = {Graph Attention Networks},
  booktitle = {International Conference on Learning Representations (ICLR)},
  year      = {2018},
}

@inproceedings{xu2019gin,
  author    = {Xu, Keyulu and Hu, Weihua and Leskovec, Jure and Jegelka, Stefanie},
  title     = {How Powerful are Graph Neural Networks?},
  booktitle = {International Conference on Learning Representations (ICLR)},
  year      = {2019},
}

@inproceedings{cao2015towards,
  author    = {Cao, Yinzhi and Yang, Junfeng},
  title     = {Towards Making Systems Forget with Machine Unlearning},
  booktitle = {IEEE Symposium on Security and Privacy},
  year      = {2015},
  pages     = {463--480},
}

@inproceedings{bourtoule2021machine,
  author    = {Bourtoule, Lucas and Chandrasekaran, Varun and Choquette-Choo, Christopher A. and Jia, Hengrui and Travers, Adelin and Zhang, Baiwu and Lie, David and Papernot, Nicolas},
  title     = {Machine Unlearning},
  booktitle = {IEEE Symposium on Security and Privacy},
  year      = {2021},
  pages     = {141--159},
}

@inproceedings{chen2022graph,
  title={Graph unlearning},
  author={Chen, Min and Zhang, Zhikun and Wang, Tianhao and Backes, Michael and Humbert, Mathias and Zhang, Yang},
  booktitle={Proceedings of the 2022 ACM SIGSAC Conference on Computer and Communications Security (CCS '22)},
  pages={499--513},
  year={2022}
}

@inproceedings{wu2023gif,
  author    = {Wu, Jiancan and Yang, Yi and Qian, Yuchun and Sui, Yongduo and Wang, Xiang and He, Xiangnan},
  title     = {{GIF}: A General Graph Unlearning Strategy via Influence Function},
  booktitle = {Proceedings of the ACM Web Conference 2023 (WWW '23)},
  year      = {2023},
  pages     = {651--661},
  doi       = {10.1145/3543507.3583521},
}

@inproceedings{dong2024idea,
  title={{IDEA}: A flexible framework of certified unlearning for graph neural networks},
  author={Dong, Yushun and Zhang, Binchi and Lei, Zhenyu and Zou, Na and Li, Jundong},
  booktitle={Proceedings of the 30th ACM SIGKDD Conference on Knowledge Discovery and Data Mining},
  pages={621--630},
  year={2024}
}

@inproceedings{cheng2023gnndelete,
  author    = {Cheng, Jiali and Dasoulas, George and He, Huan and Agarwal, Chirag and Zitnik, Marinka},
  title     = {{GNNDelete}: A General Strategy for Unlearning in Graph Neural Networks},
  booktitle = {International Conference on Learning Representations (ICLR)},
  year      = {2023},
}

@inproceedings{pan2023unlearning,
  title={Unlearning graph classifiers with limited data resources},
  author={Pan, Chao and Chien, Eli and Milenkovic, Olgica},
  booktitle={Proceedings of the ACM Web Conference 2023},
  pages={716--726},
  year={2023}
}

@article{fan2026opengu,
  title={{OpenGU}: A comprehensive benchmark for graph unlearning},
  author={Fan, Bowen and Ai, Yuming and Li, Xunkai and Guo, Zhilin and Zhu, Lei and Zeng, Guang and Li, Rong-Hua and Wang, Guoren},
  journal={Advances in Neural Information Processing Systems},
  volume={38},
  year={2026}
}

@article{said2023survey,
  title={A survey of graph unlearning},
  author={Said, Anwar and Zhao, Yuying and Derr, Tyler and Shabbir, Mudassir and Abbas, Waseem and Koutsoukos, Xenofon},
  journal={arXiv preprint arXiv:2310.02164},
  year={2023}
}

@article{mantelero2013eu,
  title={The {EU} proposal for a {General Data Protection Regulation} and the roots of the `right to be forgotten'},
  author={Mantelero, Alessandro},
  journal={Computer Law \& Security Review},
  volume={29},
  number={3},
  pages={229--235},
  year={2013},
  publisher={Elsevier}
}

@article{pardau2018ccpa,
  author  = {Pardau, Stuart L.},
  title   = {The {California Consumer Privacy Act}: Towards a {European}-Style Privacy Regime in the {United States}},
  journal = {Journal of Technology Law \& Policy},
  volume  = {23},
  year    = {2018},
  pages   = {68--114},
}

@article{xue2026towards,
  title={Towards reliable forgetting: A survey on machine unlearning verification},
  author={Xue, Lulu and Hu, Shengshan and Lu, Wei and Shen, Yan and Li, Dongxu and Guo, Peijin and Zhou, Ziqi and Li, Minghui and Zhang, Yanjun and Zhang, Leo},
  journal={ACM Computing Surveys},
  volume={58},
  number={12},
  pages={1--35},
  year={2026},
  publisher={ACM New York, NY}
}

@inproceedings{wu2021adapting,
  title={Adapting membership inference attacks to {GNN} for graph classification: Approaches and implications},
  author={Wu, Bang and Yang, Xiangwen and Pan, Shirui and Yuan, Xingliang},
  booktitle={2021 IEEE International Conference on Data Mining (ICDM)},
  pages={1421--1426},
  year={2021},
  organization={IEEE}
}

@inproceedings{liu2026a,
  title     = {A {SISA-based} Machine Unlearning Framework for Power Transformer Inter-Turn Short-Circuit Fault Localization},
  author={Liu, Nanhong and Yan, Jingyi and Sun, Mucun and Zhang, Jie},
  booktitle={2026 IEEE Power \& Energy Society General Meeting (PESGM)},
  year={2026},
}

@misc{utdhpc,
  author       = {{High Performance Computing at UTD}},
  title        = {{Home - HPC@UTD}},
  howpublished = {\url{https://hpc.utdallas.edu/}},
  note         = {accessed May 2026}
}

@article{dubey2026cost,
  title={Cost-Aware Detection and Mitigation of Charge Manipulation Attacks in Electric Vehicle Aggregations},
  author={Dubey, Rahul Kumar and Panigrahi, Bijaya Ketan},
  journal={IEEE Transactions on Smart Grid},
  year={2026},
  publisher={IEEE}
}

@article{musleh2019survey,
  title={A survey on the detection algorithms for false data injection attacks in smart grids},
  author={Musleh, Ahmed S and Chen, Guo and Dong, Zhao Yang},
  journal={IEEE Transactions on Smart Grid},
  volume={11},
  number={3},
  pages={2218--2234},
  year={2019},
  publisher={IEEE}
}

@article{caleb2026false,
  title={False Data Injection Attacks on Smart Grids: Attack Models, Challenges and Future Directions},
  author={Caleb, Toro Dama and Shao, Sicong and Kaabouch, Naima},
  journal={International Journal of Information Security},
  volume={25},
  number={3},
  pages={94},
  year={2026},
  publisher={Springer}
}

@article{boyaci2021graph,
  title={Graph neural networks based detection of stealth false data injection attacks in smart grids},
  author={Boyaci, Osman and Umunnakwe, Amarachi and Sahu, Abhijeet and Narimani, Mohammad Rasoul and Ismail, Muhammad and Davis, Katherine R and Serpedin, Erchin},
  journal={IEEE Systems Journal},
  volume={16},
  number={2},
  pages={2946--2957},
  year={2021},
  publisher={IEEE}
}

@article{xu2024task,
  title={Task-aware machine unlearning and its application in load forecasting},
  author={Xu, Wangkun and Teng, Fei},
  journal={IEEE Transactions on Power Systems},
  volume={39},
  number={6},
  pages={7178--7189},
  year={2024},
  publisher={IEEE}
}

@inproceedings{koh2017understanding,
  title={Understanding black-box predictions via influence functions},
  author={Koh, Pang Wei and Liang, Percy},
  booktitle={International Conference on Machine Learning (ICML)},
  pages={1885--1894},
  year={2017},
  organization={PMLR}
}

@article{olojede2025topology,
  title={Topology Informed Transformer for Cyber Attack Detection in Grid-Connected {PV} Systems},
  author={Olojede, Damilola R and Uddin, Md Joshem and Jacob, Roshni Anna and Coskunuzer, Baris and Zhang, Jie},
  journal={IEEE Transactions on Sustainable Energy},
  year={2025},
  publisher={IEEE}
}

@article{uddin2025mp,
  title = {{MP-Grid}: Detecting power grid outages with topological machine learning},
  author={Uddin, Md Joshem and Olojede, Damilola R and Jacob, Roshni Anna and Coskunuzer, Baris and Zhang, Jie},
  journal = {Applied Energy},
  volume = {410},
  pages = {127501},
  year = {2026},
}

@inproceedings{liu2026privacy,
  title={Privacy-Preserving Graph Unlearning for Cyberattack Detection in Electric Vehicle Charging Networks},
  author={Liu, Nanhong and Jacob, Roshni Anna and Zhang, Jie},
  booktitle={International Design Engineering Technical Conferences and Computers and Information in Engineering Conference},
  volume={},
  pages={},
  year={2026},
  organization={American Society of Mechanical Engineers}
}

@misc{eu2024aiact,
  author       = {{European Parliament and Council of the European Union}},
  title        = {Regulation ({EU}) 2024/1689 laying down harmonised rules on artificial intelligence ({Artificial Intelligence Act})},
  howpublished = {Official Journal of the European Union, OJ L, 2024/1689},
  month        = jul,
  url          = {http://data.europa.eu/eli/reg/2024/1689/oj},
  year         = {2024}
}

@inproceedings{warnecke2021machine,
  title={Machine unlearning of features and labels},
  author={Alexander Warnecke and Lukas Pirch and Christian Wressnegger and Konrad Rieck},
  year={2023},
  booktitle={Proceedings of the 30th Network and Distributed System Security (NDSS)}
}

@article{acharya2020public,
  title={Public plug-in electric vehicles+ grid data: Is a new cyberattack vector viable?},
  author={Acharya, Samrat and Dvorkin, Yury and Karri, Ramesh},
  journal={IEEE Transactions on Smart Grid},
  volume={11},
  number={6},
  pages={5099--5113},
  year={2020},
  publisher={IEEE}
}

@article{chahine2026survey,
  title={A survey of machine unlearning for electric power systems: From privacy compliance to resilient grid operations},
  author={Chahine, Khaled and El Nashef, Georges Zakka and Karim, Abdel Karim Abdel and Al Atem, Marc and Noura, Hassan N and Arnaout, Mohamad},
  journal={Array},
  pages={100987},
  year={2026},
  publisher={Elsevier}
}

@article{lee2024learning,
  title={Learning and unlearning to operate profitable secure electric vehicle charging},
  author={Lee, Sangyoon and Choi, Dae-Hyun},
  journal={IEEE Transactions on Industrial Informatics},
  volume={20},
  number={9},
  pages={11213--11223},
  year={2024},
  publisher={IEEE}
}

@article{paphitis2026machine,
  title={Machine Unlearning in Smart Grids: A Cautionary Tale of Infrastructure Vulnerability},
  author={Paphitis, George and Marnerides, Angelos K and Panteli, Mathaios},
  journal={IEEE Systems, Man, and Cybernetics Letters},
  year={2026},
  publisher={IEEE}
}

@article{poudel2026bidirectional,
  title={Bidirectional GNN-based intrusion detection of malware injection attacks in EV charging stations},
  author={Poudel, Sushil and Baugh, J Eileen and Abouyoussef, Mahmoud and Takiddin, Abdulrahman and Ismail, Muhammad and Refaat, Shady S},
  journal={IEEE Transactions on Intelligent Transportation Systems},
  year={2026},
  publisher={IEEE}
}

@inproceedings{jain2026is,
  title     = {Is Graph Unlearning Ready for Practice? A Benchmark on Efficiency, Utility, and Forgetting},
  author    = {Samyak Jain and Ronak Kalvani and Sainyam Galhotra and Sayan Ranu},
  booktitle = {The Fourteenth International Conference on Learning Representations},
  year      = {2026},
}

@article{ding2024understanding,
  title={Understanding fine-tuning in approximate unlearning: A theoretical perspective},
  author={Ding, Meng and Sharma, Rohan and Chen, Changyou and Xu, Jinhui and Ji, Kaiyi},
  journal={arXiv preprint arXiv:2410.03833},
  year={2024}
}

@article{sadric2026graph,
  title={Graph Machine Learning: An Opportunity for Power Systems},
  author={Sadric, Martin and P{\"u}tz, Sebastian and Nauck, Christian and Hagenmeyer, Veit and Hellmann, Frank and Witthaut, Dirk and Sch{\"a}fer, Benjamin},
  journal={arXiv preprint arXiv:2608.16494},
  year={2026}
}

@misc{ec2025dataact,
  author       = {{European Commission}},
  title        = {{EU} {Data Act} gives users control over data from connected devices},
  howpublished = {\url{https://digital-strategy.ec.europa.eu/en/news/eu-data-act-gives-users-control-over-data-connected-devices}},
  month        = sep,
  year         = {2025},
  note         = {accessed Oct. 2026}
}

@misc{enode2025dataact,
  author       = {{Enode}},
  title        = {The {EU} {Data Act}: What the energy industry needs to know before {September} 12},
  howpublished = {\url{https://enode.com/updates/the-eu-data-act-what-the-energy-industry-needs-to-know-before-september-12}},
  month        = aug,
  year         = {2025},
  note         = {accessed Oct. 2026}
}

@article{jia2023model,
  title={Model sparsity can simplify machine unlearning},
  author={Jia, Jinghan and Liu, Jiancheng and Ram, Parikshit and Yao, Yuguang and Liu, Gaowen and Liu, Yang and Sharma, Pranay and Liu, Sijia},
  journal={Advances in neural information processing systems},
  volume={36},
  pages={51584--51605},
  year={2023}
}

@article{kolipaka2024cognac,
  title={A cognac shot to forget bad memories: corrective unlearning for graph neural networks},
  author={Kolipaka, Varshita and Sinha, Akshit and Mishra, Debangan and Kumar, Sumit and Arun, Arvindh and Goel, Shashwat and Kumaraguru, Ponnurangam},
  journal={arXiv preprint arXiv:2412.00789},
  year={2024}
}

@inproceedings{zhang2025dynamic,
  title={Dynamic graph unlearning: a general and efficient post-processing method via gradient transformation},
  author={Zhang, He and Wu, Bang and Yang, Xiangwen and Yuan, Xingliang and Liu, Xiaoning and Yi, Xun},
  booktitle={Proceedings of the ACM on Web Conference 2025},
  pages={931--944},
  year={2025}
}

@article{wu1989network,
  title={Network reconfiguration in distribution systems for loss reduction and load balancing},
  author={Wu, Felix F and Baran, Mesut},
  journal={IEEE Transactions on Power delivery},
  year={1989}
}

@article{haider2026physics,
  title={Physics-Aware Machine Unlearning for Cyber-Physical Systems},
  author={Haider, Mohammad Zakaria and Nadeem, Muhammad and Rahman, Mohammad Ashiqur},
  journal={arXiv preprint arXiv:2609.36633},
  year={2026}
}

@article{dvorkin2020differentially,
  title={Differentially private optimal power flow for distribution grids},
  author={Dvorkin, Vladimir and Fioretto, Ferdinando and Van Hentenryck, Pascal and Pinson, Pierre and Kazempour, Jalal},
  journal={IEEE Transactions on Power Systems},
  volume={36},
  number={3},
  pages={2186--2196},
  year={2020},
  publisher={IEEE}
}

@incollection{wilcoxon1992individual,
  title={Individual comparisons by ranking methods},
  author={Wilcoxon, Frank},
  booktitle={Breakthroughs in statistics: Methodology and distribution},
  pages={196--202},
  year={1992},
  publisher={Springer}
}

\end{document}